# Gas Analyses of First complete JET Cryopump Regeneration with ITER-Like Wall

S. Grünhagen Romanelli[1], S. Brezinsek[2], B. Butler[1], J.P. Coad[3], A. Drenik[4], C. Giroud[1], S. Jachmich[5], T. Keenan[1], U. Kruezi[1], M. Mozetic[4], M. Oberkofler[6], A. Parracho[1], M. Romanelli[1], R. Smith[1] and JET EFDA contributors*

*JET-EFDA, Culham Science Centre, OX14 3DB, Abingdon, UK*

[1]EURATOM-CCFE Fusion Association, Culham Science Centre, OX14 3DB, Abingdon, OXON, UK
[2]Institut für Energie- und Klimaforschung-Plasmaphysik, Forschungszentrum Jülich, TEC, Association EURATOM-FZJ, D-52425, Jülich, Germany
[3]Association EURATOM-TEKES, VTT, PO Box 1000, 02044 VTT, Espoo, Finland
[4]Jožef Stefan Institute, Jamova 39, 1000 Ljubljana, Slovenia
[5]LPP-ERM/KMS, Association EURATOM-Belgian State, B-1000 Brussels, Belgium
[6]Association EURATOM-Max-Planck-Institut für Plasmaphysik, 85748 Garching, Germany
* See annex of F. Romanelli et al, "Overview of JET Results",
(24th IAEA Fusion Energy Conference, San Diego, USA (2012)).

**ABSTRACT.**

Analytical results of a complete JET cryopump regeneration, including the nitrogen panel, following the first ITER-Like Wall campaign are presented along with the in-situ analyses of residual gas. H/D mixtures and impurities such as nitrogen and neon were injected during plasma operation in the vessel to study radiation cooling in the scrape-off-layer and divertor region. The global gas inventory over the campaign is incomplete, suggesting residual volatile impurities are remaining on the cryogenic panel. This paper presents results on a) residual deuterium on the panel which is related to the campaign very low, b) impurities like nitrogen which sticks on the panel and c) the ammonia production which can be observed in the RGA spectrum.

## 1. INTRODUCTION

During the 2010 shutdown all JET Plasma-Facing Components (PFCs) made of Carbon- Fibre Composite (CFC) where replaced with bulk Beryllium or Be-coated tiles and Tungsten (W) tiles in the divertor to mimic the configuration envisaged for ITER, ITER-Like Wall (ILW). Following the installation a eleventh month experimental campaign was executed to investigate key issues of plasma-wall interaction such as plasma compatibility of the tungsten divertor and fuel retention in PFCs. In order to minimise the W sputtering and allow long operation on the inertially cooled PFCs, impurity seeding is used to mitigate the divertor heat load by radiation cooling. The radiation cooling can be induced by a combination of high recycling and seeding of impurities such as noble gases or nitrogen which has a radiation potential optimised for the temperatures in the divertor. A critical issue with the usage of nitrogen in a metallic environment is the formation of nitrides and ammonia or other derivatives. In JET dedicated experiments have been executed to study the impact of nitrogen on the PFCs and in particular on the production of chemical reactive species such as ammonia ($NQ_3$, $Q = H,D$) in the gas exhaust [1,2].

## 2. JET PUMPED DIVERTOR CRYOPUMP REGENERATION

The JET tokamak is equipped with a Pumped Divertor (PD) with integrated cryogenic pump. Hydrogen isotopes and impurities injected or generated during plasma operations are primarily pumped out and trapped by the cryogenic pump. The pump consists of two panels cooled by a) liquid helium (LHe) and b) liquid nitrogen ($LN_2$). The LHe panel which is trapping mostly hydrogen isotopes was frequently regenerated (78 times) at 77K for either dedicated gas balance experiments [3] or when the $D_2$ inventory limit of 30000 $Pam^3$ was reached during the campaign whereas the $LN_2$ panel only was regenerated once at the end of the experimental campaign in order not to compromise the good vessel conditions by releasing pumped water. Two tons of W and two tons of Be of new plasma-facing material were installed and no $LN_2$ regeneration after the initial backing cycle and glow discharge operation has been executed. The first and so far only complete regeneration under JET-ILW conditions representing 11 months of operation can be compared with a complete regeneration at the end of the JET-C operation, but collecting only one month of JET operation with carbon-based PFCs. First comparison between the exhaust from the two wall configurations can be made.

*2.1 LIQUID HELIUM PANEL REGENERATION*

At the end of the last day of plasma operation completing a period of 151 identical H-plasmas including two light cryogenic regenerations (LHe panel) [3] and following an out-gassing time of 15 hours for the PFCs (vacuum valve closed to Neutral beam boxes and already warmed-up to 77K), the liquid helium supply to the cryopump was isolated. The LHe panel which had had an accumulation time of one day of operation (10 discharges), gradually warmed up and the released gas which passed through the torus vessel and was mechanically turbo pumped via the horizontal pumping duct over a pipe bridge to the Active Gas Handling System (AGHS) [4]. The AGHS from its side employed in addition a mechanical pump to collect the released gas in a reservoir and performed Pressure-Volume-Temperature-concentration (PVT-c) analyses. The 2210 Pam$^3$ at 293.15K collected were analysed with two Gas Chromatographs (GC) and a Residual Gas Analyzer (RGA). Results are in vol%: 0.2% He, 0.3% H$_2$, 2.6% HD, 95.7% D$_2$, 0.9% Ar, 0.08% N$_2$, 0.14% CO$_2$, 0.09% methane, <0.03% C2 to C6. A contribution via the constant air leak would be 100% in the case of N$_2$, 0.3% for Ar and 0.1% for CO$_2$.

*2.2 LIQUID NITROGEN PANEL REGENERATION*

After the complete regenerating of the LHe panel and pumping of the remaining gas from the transfer lines, the liquid nitrogen coolant was stopped. Conventional ND radicals during N$_2$ seeding experiments were observed by divertor spectrometer system as described in [5] which could indicate ammonia formation via plasma /surface reactions [1]. No such ammonia in plasma formation but CN production by surface interaction was observed in JET-C experiments due to surface interaction with the graphite CFC [5]. Ammonia would be trapped on the LN$_2$ panel together with water which origin mostly through D$_2$ glow discharge operation (100h) but also from the initial pump down after the ILW installation and a small constant air leak (<8x10$^{-5}$ Pam$^3$/s). In order to avoid mixing-up NQ$_3$ and water and the possibility to create corrosive ammonium hydroxide which will be immediately adsorb onto surfaces in the stainless steel pipework and reservoirs, the released gas was collected in separate volumes. NQ$_3$ which has a lower boiling point than water is expected to be released first from the LN$_2$ panel. The first batch of released gas at a LN$_2$ panel-temperature from 77K to 200K was collected at AGHS on a cold trap cooled to 77K and further connected to a 25L reservoir. The cold trap was later warmed up and the gas analyzed. The second batch from a LN$_2$ panel-temperature from 200K until ambient temperature was collected via a mechanical roots pump leading to the same 25L reservoir as the first gas collection. The reservoir contained both the non condensate components of the first batch and the total of the second collection. Finally this reservoir was sampled and analysed with both the Gas Chromatography and RGA technique. A total of 830 Pam$^3$ (293.15K) of released gas was collected.

## 3. Gas Chromatograph and RGA analyses
### *3.1 METHANE AND HIGHER HYDROCARBONS*

Hydrogen isotope species, molecular nitrogen, air components, carbon oxides, methane and long-chain hydrocarbons are analysed by common standard GC technology. The total of the $LN_2$ panel in vol % is: 30% $Q_2$, 13% $N_2$, 2% $CQ_4$ and 11% long-chain hydrocarbons (HC). The sample showed a reduction and different profile of long-chain HC as compared with the JET-C (carbon-wall) data (1-month-operation) [6]. A fast decay of C in the residual plasma and of methane during the first year of ILW plasma operation was observed which indicates that hydrocarbons mostly originated from the initial operation phase [7] and levelled at about 0.1% which is the detection limit for methane, Figure 1. This indicated that a large fraction of the initially observed methane is resulting from the installation and plasma clean-up the ILW which include new material out-gassing etc.. The JET-C data were recorded long after a major installation of new wall components with a well conditioned carbon wall (>10 years). Overall the HC concentration is much lower as compared to the JET-C wall. But the HC distribution showed a different pattern compared to the carbon wall e.g. C2 to C4 is 16 times less whereas the C8 to C10 group show an increase of four times. The longest chains might be related to cleaning substrates (oil).

### *3.2 AMMONIA AND WATER ANALYSES*

The use of the existing AGHS analytical systems, GC and internal-RGA, for the analysis of ammonia and water has been assessed. GC is not the preferred analytical technique to detect ammonia and water due to their high polarity and corrosive attribute in the case of $NQ_3$. Only qualitative analyses can be performed, but an attempt has been made to calibrate the GC system with a 500ppm ammonia standard, this is the highest recommended concentration allowed to enter the GC without damaging the system by etching. However, no definite signal for $NQ_3$ could be obtained on the GC separation columns, PoraPLOT Q and CP-Sil 5CB, despite literature giving elution factors for ammonia for these columns. Also the internal-RGA connected to the same sample manifold as the GCs has shown no signal at mass-to-charge-ratio (m/z) = 17 atomic mass unit (amu). One explanation of the above observation could be that the 500ppm $NH_3$ concentration is too low and the ammonia gets entirely adsorbed onto the surfaces. Ammonia is prone to adsorb onto surfaces especially in combination with water as water exacerbates adsorption of ammonium onto internal surfaces. The ammonia standard gas cylinder was connected to the analytical system via a 2m long and ¼" in diameter stainless steel pipework. The analysis of the first gas batch collected from the $LN_2$ panel has shown a multiple, overlaid asymmetric peak where the literature gives ammonia retention, Figure 2. Water was detected by GC but could not been quantified. Another evidence for the presence of ammonia and water is that the humidity sensor installed in the sampling line gave positive indication in both gas collections for polar molecule. The unquantifiable compounds are estimated to account for in the range of 40% of the released gas of the $LN_2$ panel and could be water and ammonia.

## 3.3 RGA ANALYSES

Several RGAs installed at the JET vessel were used to measure the partial pressures of gases released from the $LN_2$ panel. One analyzer RGA1 (HIDEN) [8] is installed in the sub-divertor region, in a vertical duct (Octant 8) close to the divertor cryopump – closest to the released gas from the panels. Two further analyzers RGA2 (Pfeiffer) and RGA3 (HIDEN) are positioned in a bypass section in the main horizontal pump duct of the JET vessel in front of the torus exhaust turbo pump leading to the AGHS. A fourth analyzer, internal-RGA or RGA4 was employed in the AGHS installed in the same sample manifold leading to the GC systems. All four RGAs were operated in a range of m/z = 1–100 amu and have shown slightly different signals which is plausible considering the different instrument type and installation position. Additionally adsorption and re-adsorption from surfaces (torus vessel ~180m$^3$, pipework ~11m$^3$) during the pumping process transferring the gas to AGHS might lead to a slight variation of the RGA signals. The most dominant signal in the 16-20 amu mass range was observed on RGA1 which is the closest to the $LN_2$ panel and might indicate the presence of deuterated water and deuterated ammonia. Methane can be neglected due to low concentration (GC <2%). The signals of all employed RGAs (RGA1-4) were widely populated up to m/z= 100 amu which indicates the presence of long-chain HC and was confirmed by the GC analyses. For evaluation cracking pattern and cryogenic pump temperature has to be taken into account. A low signal on m/z= 21 amu was observed on RGA1-3 and RGA4 (1$^{st}$ gas collection), the strongest candidate is $^{15}ND_3$ following a $^{15}N_2$ seeding trace experiment at the beginning of the two weeks operation in identical H-mode plasmas before the cryopump regeneration. No other $^{15}N_2$ has been injected in the JET-ILW. A comparison between the partial pressures of m/z = 21 and m/z= 20 is considered to be an effective way to distinguish ammonia ($^{15}ND_3$) from the other species populating m/z= 20. Figure 3 shows two major peaks, one peak at 190K (m/z = 20/21) and one peak at 220K (m/z= 20). That indicates most likely the presence of ammonia (1$^{st}$ peak) and overlaid by water (2$^{nd}$ peak). Positive ammonia detection via RGA was reported from ASDEX Upgrade during $N_2$ seeding plasma operation [9].

RGA 1-4 have shown high signals on m/z = 2-4 amu ($Q_2$) and m/z = 28 amu ($N_2$, CO, C2) this was confirmed by GC analyses. $Q_2$ and $N_2$ contribute almost half to the total collected $LN_2$ panel sample. This indicates that $Q_2$ and $N_2$ which are released during frequent LHe panel regenerations might get re-trapped on the $LN_2$ surfaces possibly via cryosorption on frost formation. Therefore the $Q_2$ and $N_2$ content on the $LN_2$ panel has to be included in future gas balance calculations. The total amount of retained $D_2$ on the $LN_2$ panel is 200 Pam3 or 0.05% in comparison to the total amount of $D_2$ pumped over the campaign and light regeneration by the LHe panel (100h of $D_2$ glow discharge not included). The panel released 110 Pam$^3$ $N_2$ or 5.1% of total injected nitrogen during the campaign. In total 130 Pam$^3$ of $^{15}N_2$ or 6.1% of total nitrogen ($^{14}N_2+^{15}N_2$) was injected.

## 4. SUMMARY

Gas analyses of the first complete JET cryopump regeneration with ITER-Like Wall were performed. The released gas of the JET cryopump was collected and analysed by Gas Chromatography (GC) and RGA technique. A reduction and different profile of long-chain hydrocarbons was detected. The methane profile during the first year of ILW operation indicates that hydrocarbons mostly originated from the initial operation phase and installation of the ILW. Water was detected by GC which origin mostly through $D_2$ glow discharge operation (100h) but also from the initial pump down after the ILW installation and a small constant air leak ($<8 \times 10^{-5}$ Pam$^3$/s). The GC revealed an overlaid asymmetric peak, where literature gives ammonia retention. All employed RGAs have seen a low signal on m/z=21 amu which indicates $^{15}ND_3$ presence. The comparison between the partial pressures of m/z= 21 and m/z= 20 is an effective way to distinguish ammonia ($^{15}ND_3$) from the other species populating m/z= 20.

The $LN_2$ panel sample contained 0.05% of total injected $D_2$ during the ILW. The panel released 110 Pam$^3$ $N_2$ or 5.1% of total injected nitrogen during the campaign. A clear identification of ammonia has to be confirmed by ammonia calibration such as (i) ammonia calibration of GC and RGA at AGHS with higher $NH_3$ concentration (ii) ammonia calibration of JET vessel RGAs (iii) improved set-up and calibration of pressure gauges close to the JET cryopump and (iv) slower warm-up of the cryopump to take advantage of the different boiling points;

$$T(methane) < T(ammonia) < T(water).$$

An upgrade/calibration of the present JET analytical system is foreseen for the next experimental campaign.

**Acknowledgements**
This work, supported by the European Communities under the contract of Association between EURATOM and CCFE, was carried out within the framework of the European Fusion Development Agreement. The views and opinions expressed herein do not necessarily reflect those of the European Commission. This work was also part-funded by the RCUK Energy Programme under grant EP/I501045.

**REFERENCES**
[1]. M. Oberkofler, D. Douai, S. Brezinsek et al. *First nitrogen-seeding experiments in JET with the ITER-like Wall*, Journal of Nuclear Materials (2013), DOI: http://dx.doi.org/10.1016/j.jnucmat.2013.01.041
[2]. C. Giroud et al., *Nitrogen Seeding for Heat Load Control in JET ELMy H-mode Plasmas and its Compatibility with ILW Materials*, proceeding of 24th IAEA Fusion Energy Conference (FEC2012), San Diego, USA
[3]. S. Brezinsek, et al., *Fuel Retention Studies with the ITER-like Wall in JET*, IOP EFDA–JET–PR(12)26, Preprint of Paper to be submitted for publication in Nuclear Fusion (2012)
[4]. R. Lässer et al, "Overview of the performance of the JET Active Gas Handling System during and after DTE1", *Fusion Engineering and Design* **47**, 173-203 (1999)
[5]. S. Brezinsek et al., *Impact of nitrogen seeding on carbon erosion in the JET divertor*, Journal of Nuclear Materials **417**, 624-628, (2011)

[6] S. Grünhagen et al. *A*nalysis of hydrocarbons of the JET divertor cryogenic pump at the end of the carbon wall campaign using a micro gas chromatograph, *Fusion Science and Technology* **60**, No 3, 931-936 (2011)

[7]. S. Brezinsek et al., Residual carbon content in the initial ITER-Like Wall experiments at JET, *Journal of Nuclear Materials*, online 2013 (2013)

[8]. U. Kruezi et al. JET divertor diagnostic upgrade for neutral gas analysis, Review of Scientific Instruments **83**, 10D728 (2012); http://dx.doi.org/10.1063/1.4732175

[9]. D Neuwirth et al., *Formation of ammonia during nitrogen-seeded discharges at ASDEX Upgrade* , Plasma Physics and Controlled Fusion **54** (2012) 085008 (10pp) doi:10.1088/0741-3335/54/8/085008

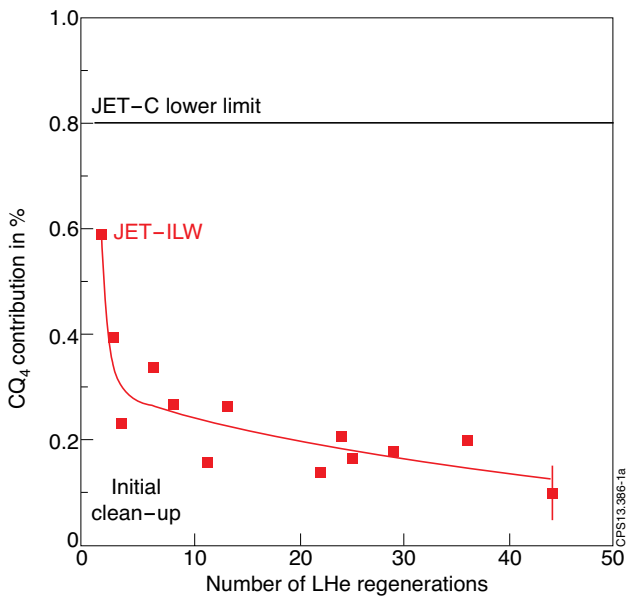

*Figure 1: Methane decrease with ILW*

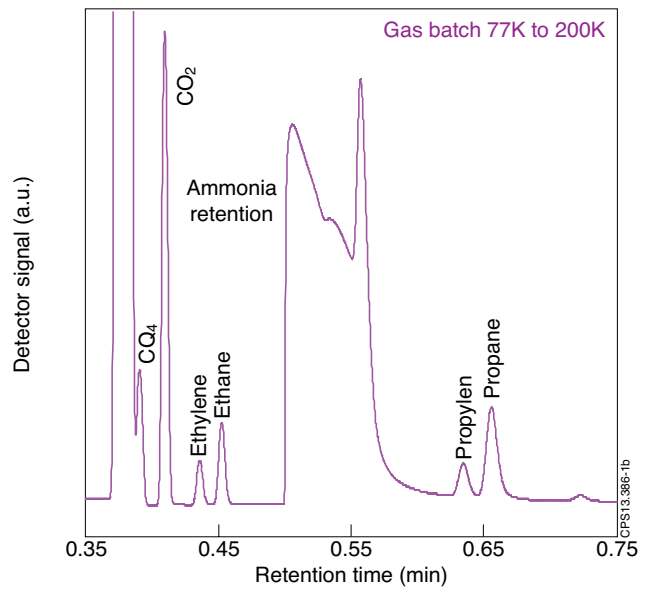

*Figure 2: GC chromatogram of 1st gas batch*

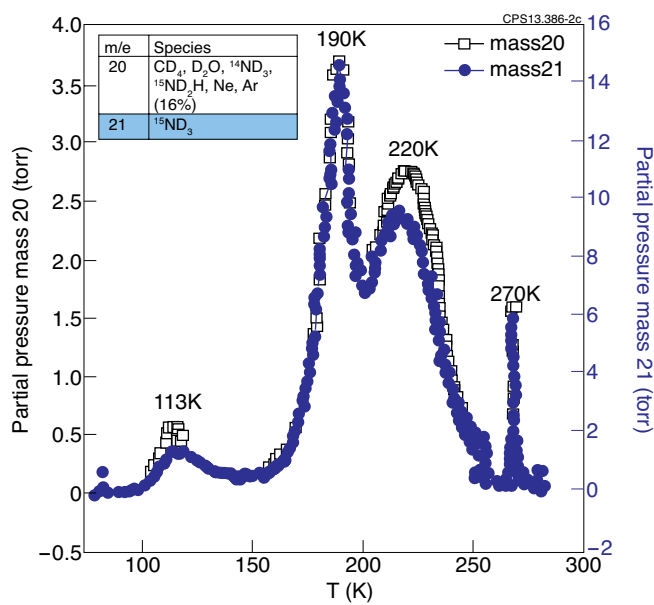

*Figure 3: RGA1, m/z= 20/21 amu*